\documentclass[aps,prl,twocolumn,showpacs,preprintnumbers,amsmath,amssymb,refmerge,superscriptaddress]{revtex4}
\usepackage{graphicx}
\usepackage{dcolumn}
\usepackage{longtable}
\newcommand{\ra}{\rightarrow}
\newcommand{\bp}{B^+}

\newcommand{\dd}{D^0\bar{D}^0}
 
\begin{document}

%\preprint{\vbox{ \hbox{   }
%                 \hbox{BELLE-CONF-0303}
%                 \hbox{Parallel Sessions: 3, 10}
%                 \hbox{EPS-ID 515}
%                 \hbox{hep-ex/0307061}
%}}

\title{ \quad\\[0.5cm]
 Observation of $B^+\ra\psi(3770)K^+$}

%                        \input{author-conf2003}
%*******************************************************************************

%%% Paper:    B+ -> psi(3770) K+
%%% Journal:  Physical Review Letters
%%% Contacts: R. Chistov (chistov@iris1.itep.ru)
%%% Non-responding authors or those who said NO are commented out.
%%% ====================================================================
%%% Click the RELOAD button on your web browser to see the updated file.
%%% ====================================================================
%%% Use \input{author} to insert this material into your latex file.
%%%%% Force institutions to appear in alphabetical order when typeset.
%%%\affiliation{Aomori University, Aomori}
\affiliation{Budker Institute of Nuclear Physics, Novosibirsk}
\affiliation{Chiba University, Chiba}
%%%\affiliation{Chuo University, Tokyo}
\affiliation{University of Cincinnati, Cincinnati, Ohio 45221}
\affiliation{University of Frankfurt, Frankfurt}
%%%\affiliation{Gyeongsang National University, Chinju}
\affiliation{University of Hawaii, Honolulu, Hawaii 96822}
\affiliation{High Energy Accelerator Research Organization (KEK), Tsukuba}
\affiliation{Hiroshima Institute of Technology, Hiroshima}
\affiliation{Institute of High Energy Physics, Chinese Academy of Sciences, Beijing}
\affiliation{Institute of High Energy Physics, Vienna}
\affiliation{Institute for Theoretical and Experimental Physics, Moscow}
\affiliation{J. Stefan Institute, Ljubljana}
\affiliation{Kanagawa University, Yokohama}
\affiliation{Korea University, Seoul}
%%%\affiliation{Kyoto University, Kyoto}
\affiliation{Kyungpook National University, Taegu}
\affiliation{Swiss Federal Institute of Technology of Lausanne, EPFL, Lausanne}
\affiliation{University of Ljubljana, Ljubljana}
\affiliation{University of Maribor, Maribor}
\affiliation{University of Melbourne, Victoria}
\affiliation{Nagoya University, Nagoya}
\affiliation{Nara Women's University, Nara}
\affiliation{National Kaohsiung Normal University, Kaohsiung}
\affiliation{National United University, Miao Li}
\affiliation{Department of Physics, National Taiwan University, Taipei}
\affiliation{H. Niewodniczanski Institute of Nuclear Physics, Krakow}
\affiliation{Nihon Dental College, Niigata}
\affiliation{Niigata University, Niigata}
\affiliation{Osaka City University, Osaka}
\affiliation{Osaka University, Osaka}
\affiliation{Panjab University, Chandigarh}
\affiliation{Peking University, Beijing}
\affiliation{Princeton University, Princeton, New Jersey 08545}
\affiliation{RIKEN BNL Research Center, Upton, New York 11973}
\affiliation{Saga University, Saga}
\affiliation{University of Science and Technology of China, Hefei}
\affiliation{Seoul National University, Seoul}
\affiliation{Sungkyunkwan University, Suwon}
\affiliation{University of Sydney, Sydney NSW}
\affiliation{Tata Institute of Fundamental Research, Bombay}
\affiliation{Toho University, Funabashi}
\affiliation{Tohoku Gakuin University, Tagajo}
\affiliation{Tohoku University, Sendai}
\affiliation{Department of Physics, University of Tokyo, Tokyo}
%%%\affiliation{Tokyo Institute of Technology, Tokyo}
\affiliation{Tokyo Metropolitan University, Tokyo}
\affiliation{Tokyo University of Agriculture and Technology, Tokyo}
%%%\affiliation{Toyama National College of Maritime Technology, Toyama}
\affiliation{University of Tsukuba, Tsukuba}
\affiliation{Utkal University, Bhubaneswer}
\affiliation{Virginia Polytechnic Institute and State University, Blacksburg, Virginia 24061}
\affiliation{Yokkaichi University, Yokkaichi}
\affiliation{Yonsei University, Seoul}
  \author{R.~Chistov}\affiliation{Institute for Theoretical and Experimental Physics, Moscow} % ITEP
  \author{K.~Abe}\affiliation{High Energy Accelerator Research Organization (KEK), Tsukuba} % KEK
  \author{K.~Abe}\affiliation{Tohoku Gakuin University, Tagajo} % TohokuGakuin
% \author{N.~Abe}\affiliation{Tokyo Institute of Technology, Tokyo} % TIT
  \author{T.~Abe}\affiliation{High Energy Accelerator Research Organization (KEK), Tsukuba} % KEK
% \author{I.~Adachi}\affiliation{High Energy Accelerator Research Organization (KEK), Tsukuba} % KEK
  \author{H.~Aihara}\affiliation{Department of Physics, University of Tokyo, Tokyo} % Tokyo
  \author{M.~Akatsu}\affiliation{Nagoya University, Nagoya} % Nagoya
% \author{M.~Asai}\affiliation{Hiroshima Institute of Technology, Hiroshima} % Hiroshima
  \author{Y.~Asano}\affiliation{University of Tsukuba, Tsukuba} % Tsukuba
% \author{T.~Aso}\affiliation{Toyama National College of Maritime Technology, Toyama} % Toyama
% \author{V.~Aulchenko}\affiliation{Budker Institute of Nuclear Physics, Novosibirsk} % BINP
  \author{T.~Aushev}\affiliation{Institute for Theoretical and Experimental Physics, Moscow} % ITEP
% \author{S.~Bahinipati}\affiliation{University of Cincinnati, Cincinnati, Ohio 45221} % Cincinnati
  \author{A.~M.~Bakich}\affiliation{University of Sydney, Sydney NSW} % Sydney
  \author{Y.~Ban}\affiliation{Peking University, Beijing} % Peking
% \author{E.~Banas}\affiliation{H. Niewodniczanski Institute of Nuclear Physics, Krakow} % Krakow
% ...
\collaboration{The Belle Collaboration}

\begin{abstract}

%DRM We report the first observation of the decay $B^+\ra\psi(3770)K^+$
%DRM where $\psi(3770)$ is reconstructed in $\dd$ and $D^+D^-$ decay channels. 
%DRM The measured branching fraction is 
%DRM ${\cal B}(B^+\ra\psi(3770) K^+)=(0.48\pm 0.11\pm 0.12)\times10^{-3}$. 
%DRM For the decay $B^+ \to \dd K^+$ we have
%DRM measured its branching fraction to be equal to $(1.17\pm 0.21\pm
%DRM 0.25)\times 10^{-3}$. For the branching fraction of the $B^+\ra
%DRM D^+D^-K^+$ decay we have set an upper limit of $0.79\times 10^{-3}$ at
%DRM 90$\%$ confidence level.  The analysis is based on
%DRM $88\,\mathrm{fb}^{-1}$ of data collected at $\Upsilon(4S)$ resonance
%DRM by the Belle detector at the KEKB asymmetric-energy $e^+e^-$ collider.

We report the first observation of the decay $B^+\ra\psi(3770)K^+$
where $\psi(3770)$ is reconstructed in the $\dd$ and $D^+D^-$ decay channels. 
The 
%measured 
obtained
branching fraction is 
${\cal B}(B^+\ra\psi(3770) K^+)=(0.48\pm 0.11\pm 0.07)\times10^{-3}$. 
We have measured the branching fraction for the decay $B^+ \to \dd K^+$ 
to be $(1.17\pm 0.21\pm
0.15)\times 10^{-3}$.  We have set a $90\%$ confidence level  upper limit 
of $0.90\times 10^{-3}$ for the decay $B^+ \ra D^+D^-K^+$. 
We also present the results of a search for possible decays to $D\bar{D}$ and $\dd\pi^0$ 
%$B^+\ra X(3872) K^+$, $X(3872)\ra D\bar{D}^0$ 
of the recently observed $X(3872)$ particle. 
The analysis is based on
$88\,\mathrm{fb}^{-1}$ of data collected at the $\Upsilon(4S)$ resonance
by the Belle detector at the KEKB asymmetric-energy $e^+e^-$ collider.

\end{abstract}


\maketitle

{\renewcommand{\thefootnote}{\fnsymbol{footnote}}}
\setcounter{footnote}{0}

%DRM The $B$ decay modes with a charmonium in the final state are
%DRM extensively used by the Belle and BaBar providing the measurement of CP
%DRM violation parameter sin$2\phi_1$\cite{sin2phi1_1}, \cite{sin2phi1_2},
$B$ decay modes with charmonium in the final state are
%SLOextensively used by the Belle and BaBar Collaborations in their measurements of the CP
extensively used by the Belle and BaBar Collaborations for measurements of the $\mathit{CP}$
violation parameter 
sin$2\phi_1$~\cite{sin2phi1_1, sin2phi1_3}. 
Belle has recently reported the first observations of
%DRM $B^+\ra\chi_{c0}K^+$~\cite{hic0} and $B\ra\chi_{c2} {\rm X}$~\cite{hic2}
%DRM decays. The decay rates for these modes were measured to be comparable
the decays $B^+\ra\chi_{c0}K^+$~\cite{hic0} and $B\ra\chi_{c2} {X}$~\cite{hic2}. 
The decay rates for these modes were measured to be comparable
to those for $J/\psi$ and $\psi(2S)$. 

%%DRM I believe that by ``known'' you mean the charmonia already
%%DRM seen in B decays.  Of course, these states are ``known'' in 
%%DRM other ways.
%DRM In contrast to the known charmonia seen so far in $B$ decays, the
In contrast to the charmonia seen so far in $B$ decays, the
$\psi(3770)$ state 
% ($1^3D_1$, $J^{PC}=1^{--}$)~\cite{pdg} 
%DRM is just above open charm threshold and decays dominantly to the pair of $D$ mesons.
is just above open charm threshold and decays dominantly to pairs of $D$ mesons~\cite{pdg}.
%DRM Since the discovery of $\psi(3770)$
%DRM it is considered as the only candidate to the $D$-wave charmonium state $1^3D_1$.
%
%Since its discovery, the $\psi(3770)$
%%SLO has been considered to be the only candidate for the ${\rm D}$-wave charmonium state $1^3{\rm D}_1$.
%has been considered to be the  $1^3{\rm D}_1$, the only candidate for a ${\rm D}$-wave charmonium state.
%%DRM However its leptonic width provides an insight that   
%Its leptonic width, however, suggests that the 
%$\psi(3770)$ has an admixture of S-wave states such as 
%$\psi(2S)$~\cite{SDmixing}.  
%
The $\psi(3770)$ is generally considered to be predominantly the $1^3D_1$ charmonium state.
However, it has a non-zero leptonic width, which indicates that there is some mixing with 
the nearby $\psi(2S)$ S-wave state~\cite{SDmixing}.
%DRM The $S$-$D$-wave mixing opens for $\psi(3770)$ non-$D{\overline D}$ decay channels 
%opens non-$D{\overline D}$ decay channels for $\psi(3770)$,  
%with branching fractions depending on mixing angle.
%DRM Also the large mixing angle results in $B$ decays to     
%DRM $\psi(3770)$ and $\psi(2S)$ with the rates of the same order.  
%Moreover, 
A large ${\rm S}$-${\rm D}$-wave  
mixing angle could result in comparable decay rates for
$B$ decays to  the $\psi(3770)$ and the $\psi(2S)$. 
%DRM Independently of the $S$-$D$-mixing, for pure $D$-wave state the estimation of 
%Independent of the ${\rm S}$-${\rm D}$-mixing, 
For 
%a 
%SLO 
%pure ${\rm D}$-wave state, estimation of 
a pure ${\rm D}$-wave state, an estimate of 
${\cal B}(B\ra\psi(3770) X)$ based on the color-octet model 
%DRM gives the value of $0.28\%$~~\cite{B_to_psi3770___theoretical_predictions} 
%DRM as large as measured for $J/\psi$ and $\psi(2S)$.  
gives a value of $0.28\%$~~\cite{B_to_psi3770___theoretical_predictions}, 
which is as large as the measured values for $J/\psi$ and $\psi(2S)$.  
% The production of the $D$-wave charmonia in $B$ decays is predicted to
% be dominated by color-octet mechanism which is estimated to be 50
% times larger than color-singlet contribution. In particular, 
%DRM The experimental study of $\psi(3770)$ production in $B$ decays 
%DRM could tests theoretical models and provides additional information on the structure of 
%DRM $\psi(3770)$ wave-function.
%%DRM I am not sure that you need this sentence.  It basically just
%%DRM summarizes what has just been said.  If there is a shortage of
%%DRM space latter on, it could be omitted.  
Experimental studies of $\psi(3770)$ production in $B$ decays 
test theoretical models and provide additional information 
on the structure of the $\psi(3770)$ wave-function.  

%%DRM This paragraph also has a big overlap with the abstract,
%%DRM and can be deleted or shortened if space is needed.
%DRM In this paper, we performed a search for $\psi(3770)$ production in
%DRM $B$ decays and report the first observation of the $B^+ \to
%DRM \psi(3770)K^+$ decay.  We also studied the $B^+\ra\dd K^+$ and
In this paper, we 
%describe a search for $\psi(3770)$ production in $B$ decays and 
report the first observation of the decay $B^+ \to\psi(3770)K^+$~\cite{cc}.  
%SLO 
%\psi(3770)K^+$.  
%We also studied the $B^+\ra\dd K^+$ and
We also report measurements of the $B^+\ra\dd K^+$ and
$B^+\ra D^+D^-K^+$ decay modes~\cite{babar_and_other_previous_studies}
and searches for the decays $B^+\ra X(3872) K^+$, $X(3872)\ra D\bar{D}$ ($\dd\pi^0$).
The analysis is performed using data collected 
with the Belle detector at the KEKB asymmetric-energy $e^+e^-$ 
collider~\cite{kekb}.  The data sample consists of $88$~fb$^{-1}$ taken at the
$\Upsilon(4S)$ resonance, which corresponds to $96\times 10^6$ $B\bar B$ pairs.

%DRM The Belle detector is a general purpose spectrometer with a 1.5 T
The Belle detector is a general-purpose spectrometer with a 1.5-T
superconducting solenoid.  Charged particle tracking is performed by a
silicon vertex detector (SVD) composed of three concentric layers of
%DRM double sided silicon strip detectors, and a 50 layer drift chamber
double sided silicon strip detectors, and a 50-layer drift chamber
(CDC).  Particle identification for charged hadrons is based on the
%DRM combination of energy loss measurements $(dE/dx)$ in the CDC, time of
%DRM flight measurements (TOF) and aerogel {\v C}erenkov counter (ACC)
combination of energy-loss measurements $(dE/dx)$ in the CDC, time-of-flight
measurements (TOF) and aerogel Cherenkov counter (ACC)
information.  A CsI(Tl) electromagnetic calorimeter (ECL) located
inside the solenoid coil is used for the detection and identification
of photons and electrons.  The detection system for $K^0_L$ mesons and
%SLO muons (KLM) consists of alternating layers of resistive plate counters
muons consists of alternating layers of resistive plate counters
%DRM and 4.7~cm thick steel plates.  The Belle detector is described in
and 4.7-cm-thick steel plates.  The Belle detector is described in
detail elsewhere~\cite{belle_detector}.

%DRM We select charged pions and kaons which originate from the region
We select charged pions and kaons that originate from the region
$dr<1$~cm, $|dz|<3$~cm, where $dr$ and $dz$ are the distances of
%DRM track closest approach to the Interaction Point in the plane
closest approach to the interaction point in the plane
perpendicular to the beam axis and along the beam direction, respectively.
Charged kaons are required to satisfy 
%pass the identification criterion of
%DRM ${\cal L}(K)/({\cal L}(K)+{\cal L}(\pi))>0.6$. ${\cal L}(K/\pi)$ is a
%DRM particle identification likelihoods for the $K/\pi$ hypotheses
%SLO ${\cal L}(K)/({\cal L}(K)+{\cal L}(\pi))>0.6$. ${\cal L}(K/\pi)$ is the 
${\cal L}(K)/({\cal L}(K)+{\cal L}(\pi))>0.6$, where ${\cal L}(K/\pi)$ is the 
particle identification likelihood for the $K/\pi$ hypotheses
calculated by combining information from the TOF, ACC and $dE/dx$
measurements in the CDC.  
%A pair of the ECL clusters not associated with
%charged tracks with invariant mass within $\pm 15$~MeV$/c^2$ of the
%nominal $\pi^0$ mass is considered as a $\pi^0$ candidate.  
%SLO Neutral $\pi^0$ candidates are selected by finding
% all pairs of non-charged-track-associated ECL
% clusters having invariant mass within
% $\pm 15$~MeV/c$^2$ of the nominal $\pi^0$ mass.
% The energy of the photons from $\pi^0$'s is required to be greater than 50~MeV.
% We also require the $\pi^0$ momentum in the center of mass system (cms)
% to be greater than 0.15~GeV$/c$ to reduce combinatorial background.
Candidate $\pi^0$ mesons are identified as
pairs of non-charged-track-associated ECL
clusters that have an invariant mass within
$\pm 15$~MeV/c$^2$ of the $\pi^0$ mass.
To reduce combinatorial background, the energy of each photon
is required to be greater than 50~MeV and
the momentum of the $\pi^0$ in the center of mass system (cms) is required to be 
greater than 0.15~GeV$/c$.

%%DRM resume here 

\begin{table*}[t]
\begin{ruledtabular}
\caption {Summary of the fit results, efficiencies, statistical
significance and branching fractions for $\bp\ra D^0\bar D^0 K^+$ and
$B^+\ra D^+D^- K^+$ decays. }
\begin{tabular}{ccccc}
Mode  & $\Delta E$ yield & Efficiency($10^{-4}$) & ${\cal B}(10^{-3})$ & Significance \\
\hline 
$\bp\ra D^0\bar D^0 K^+$ & $97.5\pm 17.6$ & 8.7 & $1.17\pm 0.21\pm 0.15$ & 5.5~$\sigma$ \\
\hline 
$\bp\ra D^+D^- K^+$ & $20.7\pm 9.9$ & 5.0 & $<0.90~ (90\% ~CL)$($0.43\pm 0.21\pm 0.06$) & 2.7~$\sigma$ \\
\end{tabular}			
\label{sumD}
\end{ruledtabular}
\end{table*}

%SLO 
%%DRM The $D^0$~\cite{cc} meson is reconstructed in $K^-\pi^+$,
% The $D^0$ meson is reconstructed in the $K^-\pi^+$,
% $K^-\pi^+\pi^+\pi^-$ and $K^-\pi^+\pi^0$ modes~\cite{cc}.  The $D^+$
%%DRM meson is reconstructed through $K^-\pi^+\pi^+$ and $K^+K^-\pi^+$
%meson is reconstructed through the $K^-\pi^+\pi^+$ and $K^+K^-\pi^+$
%DRM modes.  We select $\pm 10\,\mathrm{MeV}/c^2$ $D$ signal window for
The $D^0$ meson is reconstructed in the $K^-\pi^+$,
$K^-\pi^+\pi^+\pi^-$ and $K^-\pi^+\pi^0$ modes, and the $D^+$
in the the $K^-\pi^+\pi^+$ and $K^+K^-\pi^+$
modes.
%~~\cite{cc}.  
We use a $\pm 10\,\mathrm{MeV}/c^2$ $D$ signal window for the
charged modes ($\sim 2.5\,\sigma$) and $\pm 15\,\mathrm{MeV}/c^2$ for $K^- \pi^+ \pi^0$
mode ($\sim 2\,\sigma$). 
%The mass-vertex constraint  
%SLO Mass and vertex constrained fits are applied to all $D$ candidates to improve their
Mass- and vertex-constrained fits are applied to all $D$ candidates to improve their
momentum resolution. The $B^+$ candidates ({\emph {i.e.} $D\bar{D}$ 
pairs combined with the positive kaons in the event) are
identified by their 
%center of mass system 
cms energy difference,
$\Delta E=\Sigma_i E_i - E_{\rm beam}$, and their beam-energy constrained mass,
$M_{\rm bc}=\sqrt{E^2_{\rm beam}-(\Sigma_i \vec{p}_i)^2}$, where $E_{\rm beam}=\sqrt{s}/2$
is the beam energy in the cms and $\vec{p}_i$ and $E_i$ are
%DRM the three-momenta and energies of the $B^+$ candidate decay products.
the three-momenta and energies of the $B^+$ candidate's decay products.
%We select events with $M_{bc}>5.2$~GeV$/c^2$ and $|\Delta E|<0.2$~GeV.
%SLO For the $B^+\ra\dd K^+$ we require that one $D^0$ 
%should always be reconstructed in the cleanest $D^0\ra K^-\pi^+$ mode. 
For the $B^+\ra\dd K^+$ final state, we require that one $D^0$ 
is reconstructed in the $D^0\ra K^-\pi^+$ mode, which has the smallest background. 
We accept $B$ candidates with 
$5.272$~GeV/c$^2$ $<M_{\rm bc}<5.288$~GeV$/c^2$ and $|\Delta E|<0.2$~GeV. 
%DRM To suppress the continuum background we require the ratio of second to
To suppress the continuum background we require the ratio of the second to the
zeroth Fox-Wolfram moments~\cite{fox2} $R_2$ to be less than 0.5 and 
$|{\rm cos}\theta_{{\rm thr.}}|<0.8$, where $\theta_{{\rm thr.}}$ is 
the angle between the thrust axis of the $B$ candidate and 
the thrust axis of the rest of the event.
%For the cleanest sample where both $D^0$ are reconstructed 
%in the $K\pi$ mode, the $|{\rm cos}\theta_{thr.}|<0.8$ cut is not applyied.  
The last cut is not applied for the cleanest subset of
$B$ candidates where both $D^0$'s are reconstructed in the $K\pi$ mode.
%mode where both $D^0$ are reconstructed in the $K\pi$ mode.
%DRM In case of multiple $B$ candidates we choose the only 
In the case of multiple $B$ candidates, we choose the
candidate with the smallest value of
$\chi^2=((M_{\rm bc} - M_{B^+})/\sigma_{M_{\rm bc}})^2$.

The $\Delta E$ distributions for the $B^+\ra\dd K^+$ and
$B^+\ra D^+D^- K^+$ candidates are shown
%presented 
in Fig.~\ref{dembc2},
where the superimposed curves are the results of the fits. The fit to
%DRM the $\Delta E$ distribution is performed by a sum of Gaussian with the
%DRM width fixed from Monte Carlo (MC) 
the $\Delta E$ distribution is a sum of a Gaussian with a
fixed width taken from Monte Carlo (MC) 
to describe the signal and a first order polynomial 
to parameterize the background~\cite{de_width}.  
In the fit to the $\Delta E$ distribution, the region $\Delta E<-0.08$~GeV is excluded 
to avoid contributions from other $B\ra D^{(*)}\bar{D}^{(*)}K$ decays.  
Table~\ref{sumD} summarizes the results of the fits, the
reconstruction efficiencies~\cite{efficiencies_including_intermediate_br},  
% RUSLAN including intermediate branching fractions, 
statistical significance~\cite{stat_signif} of the signals and the calculated
%DRM branching fractions, where  
branching fractions.
For the latter, 
we assume $N(B^0\bar{B^0})=N(B^+B^-)$.
%SLO For the $D^+D^-K^+$ final state significant signal is not seen and we set a $90\%$ confidence 
% upper limit. 
For the $D^+D^-K^+$ final state, a substantial signal is not seen and we set a $90\%$ confidence 
upper limit. 
%DRM The systematic error in branching fraction mesurement 
The systematic error in the branching fraction measurement 
%DRM is dominated by the uncertainty in tracking
is dominated by the uncertainty in the tracking
efficiency ($1\%$ per track), kaon identification efficiency ($2\%$
for each kaon), $\pi^0$ reconstruction efficiency ($6\%$),
$D^0$ branching fraction uncertainty (in total $8\%$), MC
%DRM statistics ($3\%$) and signal and background parameterization ($5\%$).
statistics ($3\%$), the signal and background parameterization ($5\%$).

\begin{table*}[t]
\begin{ruledtabular}
\caption { Result of the measurement of 
%$M(\psi(3770))$ and 
$\Delta m=M(\psi(3770))-M(\psi(2S))$ obtained in this paper and previous measurements.    }
\medskip
\begin{tabular}{ccccc}
 & Belle & MARK I  & DELCO  & MARK II \\
\hline
%$M(\psi(3770))$, ~MeV/c$^2$ & $3778.4\pm 3.0\pm 0.8$  & $3772\pm 6$  & $3770\pm 6$ & $3764\pm 5$ \\
%\hline
$\Delta m=M(\psi(3770))-M(\psi(2S))$, ~MeV/c$^2$ & $92.4\pm 3.0\pm 1.3$  & $88\pm 3$  & $86\pm 2$ & $80\pm 2$ \\
\end{tabular}
\label{DM}
\end{ruledtabular}
\end{table*}

\begin{figure}[t]
\centering
\includegraphics[width=0.5\textwidth]{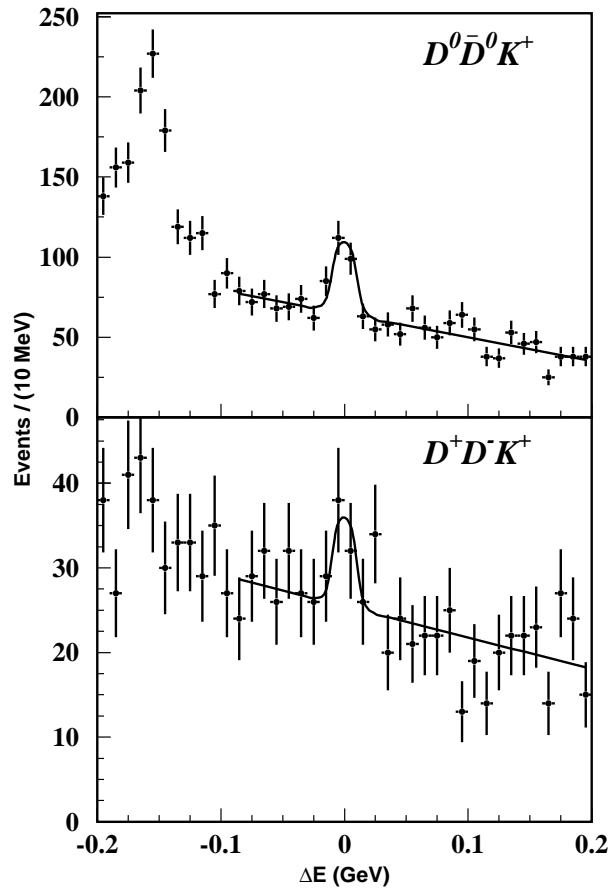}
\caption{ The $\Delta E$ distributions for
the $B^+\ra\dd K^+$(upper) and $B^+\ra D^+D^- K^+$(lower) candidates. 
Points with errors represent the data and curves show the results of the fits
described in the text.}
\label{dembc2}
\end{figure}

%SLO% We look at the $D^0\bar{D^0}$ and $D^+D^-$ invariant mass distributions from
% the $B$ signal region defined as $5.272<M_{bc}<5.288$~GeV$/c^2$ and $|\Delta
% E|<0.02$~GeV. 
% For $B^+ \to \dd K^+$

%Fig.~\ref{fitA2}(a)
We plot the $D^0\bar{D^0}$ and $D^+D^-$ invariant mass distributions for events in
the $B$ signal region defined as $5.272$~GeV/c$^2$ $<M_{bc}<5.288$~GeV$/c^2$ and $|\Delta
E|<0.02$~GeV in Fig.~\ref{fitA2}(a) and Fig.~\ref{fitA2}(b), respectively.
Here, for $B^+ \to \dd K^+$
candidates, when one of the $D^0$'s is reconstructed in the $K^-\pi^+\pi^0$
%DRM mode, we use looser $\Delta E$ cut ($|\Delta E|<0.025$~GeV) to take into
mode, we use a looser $\Delta E$ cut ($|\Delta E|<0.025$~GeV) to take into
%DRM account the worse energy resolution.  In case of multiple $B$
%SLO account the poorer energy resolution.  In the case of multiple $B$
account the poorer energy resolution due to shower leakage in the ECL.  In the case of multiple $B$
%DRM candidates we choose the only candidate which has the minimal value of
candidates, we choose the candidate with the smallest value of 
$\chi^2=(\Delta E/\sigma_{\Delta E})^2 + ((M_{\rm bc} - M_{B^+}) /
\sigma_{M_{\rm bc}})^2$.  
The $M(\dd)$ distribution has a peak at low masses, which we attribute to the $\psi(3770)$ signal.

%SLO Fig.~\ref{fitA2}(a) shows the $M(\dd)$ 
% spectrum after the described
% %DRM selection applied. The superimposed hatched histogram shows the
% $M(\dd)$ mass distribution from the $\Delta E$ sidebands~\cite{de_sdb}.
The superimposed hatched histogram in Fig.~\ref{fitA2}(a) shows the
$M(\dd)$ mass distribution for events in the $\Delta E$ sidebands~\cite{de_sdb}.
%SLO The curve (shown as solid line) is the result of the fit. The signal
% is described by relativistic Breit-Wigner
The curve (shown as solid line) is the result of a fit where the
low-mass peak is described by a relativistic p-wave Breit-Wigner
%DRM function~\cite{explanations} with free mass and natural width fixed to
function~\cite{explanations} with a floating mass and its natural width fixed to
its nominal value of $\Gamma(\psi(3770))=23.6$~MeV/c$^2$~\cite{pdg}.  
%The background is represented by the dashed line.  
%It was described by the
%The background is represented by the dashed line.  
The combinatorial background along with the contribution from non-resonant 
$B^+\ra D^0\bar{D}^0 K^+$ decays 
%SLO were described by the
% product of a two-body left threshold 
% %function (${\rm a_1}\cdot\sqrt{x-{\rm 3.729}}$) 
%and a right threshold.
% %function
% %($(4.782- x)^{{\rm a_2}}$), where $a_1$ and $a_2$ are free parameters. 
% This function is represented by the dashed line.
is described by the product of left and right two-body threshold functions
and is represented as a dashed line in Fig.~\ref{fitA2}(a).

%DRM Fit finds $\psi(3770)$ signal yield to be $N=33.6\pm 8.3$ events with
%DRM the statistical significance of $5.9\sigma$.  The mass of the
%SLO The fit yields to $\psi(3770)$ signal of $N=33.6\pm 8.3$ events with
The fit yields a $\psi(3770)$ signal of $N=33.6\pm 8.3$ events with
a statistical significance of $5.9\sigma$.  The mass of the
$\psi(3770)$ is found to be $M(\psi(3770))=3778.4\pm 3.0\pm 1.3$~MeV/c$^2$, 
%SLO 0.8$~MeV/c$^2$. From this the mass difference $\Delta
% %DRM m=M(\psi(3770))-M(\psi(2S))$ was extracted to be $92.4\pm 3.0\pm
% m=M(\psi(3770))-M(\psi(2S))$ was determined to be $92.4\pm 3.0\pm
% 0.8$~MeV/c$^2$, where we used $M(\psi(2S))=3685.96\pm
% 0.09$~MeV/c$^2$~\cite{pdg}.  Table~\ref{DM} compares our measurements of
which corresponds to a mass difference $\Delta
m=M(\psi(3770))-M(\psi(2S)) = 92.4\pm 3.0\pm
1.3$~MeV/c$^2$, where we used $M(\psi(2S))=3685.96\pm
0.09$~MeV/c$^2$~\cite{pdg}.  Table~\ref{DM} compares our measurement of
%$M(\psi(3770))$ and 
$\Delta m$ with the available results from the
%SLO MARK I, DELCO and MARK II Collaborations~\cite{pdg}, ~\cite{dm}.  We
% conclude that our measurements of $M(\psi(3770))$ 
% and $\Delta m=M(\psi(3770))-M(\psi(2S))$
%are in agreement with the results of the MARK I and DELCO
% Collaborations and $\sim 3\sigma$ above the MARK II result.  The systematic
MARK I, DELCO and MARK II Collaborations~\cite{pdg}, ~\cite{dm}; 
our measurements agree with the MARK I and DELCO results
and are $\sim 3\sigma$ above the MARK II result.  The systematic
error on the mass measurement is evaluated by varying the background function, 
%DRM the width of $\psi(3770)$ within its errors~\cite{pdg}, changing the fit range
%DRM and the bin width.
the width of the $\psi(3770)$ within its errors~\cite{pdg}, changing the fit range
and changing the bin width. It also includes the uncertainty in the $D^0$ mass. 

%The result of a fit to the $M(D^+D^-)$ distribution is given 
%by Fig.~\ref{fitA2}(b). 
%%DRM The mass for $\psi(3770)$ was fixed to the value found from the $M(\dd)$
%%DRM fit of $3778.4$~MeV$/c^2$.  Fit finds $N=7.7\pm 4.2$ events for
%The $\psi(3770)$ mass was fixed to $3778.4$~MeV/c$^2$, the value found from the $M(\dd)$ fit.
%The fit  finds $N=7.7\pm 4.2$ events for
%$B^+\ra\psi(3770)K^+$ followed by $\psi(3770)\ra D^+D^-$ .
 
%SLO From a fit to the $M(D^+D^-)$ distribution (Fig.~\ref{fitA2}(b)), we obtained a yield
% of $7.7\pm 4.2$ events. The $\psi(3770)$ mass was fixed to $3778.4$~MeV/c$^2$,
% the value found from the $M(D^0\bar{D}^{0})$ fit.
A fit to the $M(D^+D^-)$ distribution of Fig.~\ref{fitA2}(b)  yields
$7.7\pm 4.2$ events. Here the $\psi(3770)$ mass was fixed at $3778.4$~MeV/c$^2$,
the value found from the $M(D^0\bar{D}^{0})$ fit.

We studied a MC sample of generic $B\bar B$ events that 
%corresponds to
%DRM about 1.0 times the data statistics and found that the $D{\overline
%a sample the size of the data.   
%about 
has the same size as 
the data sample.
%SLO We found that the $D\bar{D}$ invariant mass exhibits smooth behavior without peaks.  We also
We find that the $D\bar{D}$ invariant mass exhibits a smooth behavior without peaks.  We also
analyzed off-resonance data taken 60~MeV below the $\Upsilon(4S)$ with
%DRM the statistics of $10$~fb$^{-1}$.  Same selection of $\dd K^+$
%DRM combinations results in one event at the whole $M(\dd)$ region that
%DRM corresponds to a negligible contribution from the continuum events.
%SLO the statistics of $10$~fb$^{-1}$.  The same selection applied to $\dd K^+$
a $10$~fb$^{-1}$ data sample.  The same selection applied to $\dd K^+$
combinations results in one event over the whole $M(\dd)$ region, which 
corresponds to a negligible contribution from the continuum. 

%%DRM I have added a paragraph break here since this is new topic.

%DRM We also studied the helicity distribution for
%DRM $B^+\ra\psi(3770) K^+$.  We fitted the $M(\dd)$ and found the
%DRM $\psi(3770)\ra\dd$ yield in each of eight ${\rm
%SLO We studied the helicity distribution for
% $B^+\ra\psi(3770) K^+$ by fitting the  $M(\dd)$ distribution 
% to determine the $\psi(3770)\ra\dd$ yield in each of eight 
% ${\rm cos}\theta_{\psi(3770)}$ bins~\cite{def_helicity}.  
% The corresponding distribution is given by Fig.~\ref{helicity}. 
The $\psi(3770)\ra\dd$ helicity distribution, determined
by fitting the  $M(\dd)$ distribution for the $\psi(3770)$ yield in each of eight 
${\rm cos}\theta_{\psi(3770)}$ bins~\cite{def_helicity},  
is shown in Fig.~\ref{helicity}. 
%%DRM please check this carefully, since I am not completely sure that
%%DRM I have not changed the meaning.   
%DRM The points are data and the histogram gives the result of the fit 
%DRM by the expected form found from the signal MC. 
% RUSLAN   The points are data and the histogram gives the result 
% RUSLAN   expected from the signal MC.  
The points are data and the histogram gives the result of a fit 
%SLO by the expected form obtained from the signal MC.
using MC-based expectations for a $J^{PC} = 1^{--}~\psi(3770)$ with floated normalization.
%DRM Since $\psi(3770)$ has $J^{PC}=1^{--}$~\cite{pdg} the   
The confidence level
%probability 
of the fit is $10.5\%$.
%SLO Since the $\psi(3770)$ has $J^{PC}=1^{--}$~\cite{pdg} the   
% $B^+\ra\psi(3770)K^+$ was modeled as a $P\ra VP$ decay followed by $V\ra PP$. 
%Fig.~\ref{helicity} (b)) presents the helicity 
%distribution for $B^+\ra\psi(3770) K^+$ followed by $\psi(3770)\ra D^+D^-$. 
%%DRM Here, due to the limited statistics we took events from the region of $M(D^+D^-)<3.8$~GeV/c$^2$.
%Here, due to the limited statistics we took events from the region $M(D^+D^-)<3800$~MeV/c$^2$.

\begin{figure}[h]
\centering
\includegraphics[width=0.5\textwidth]{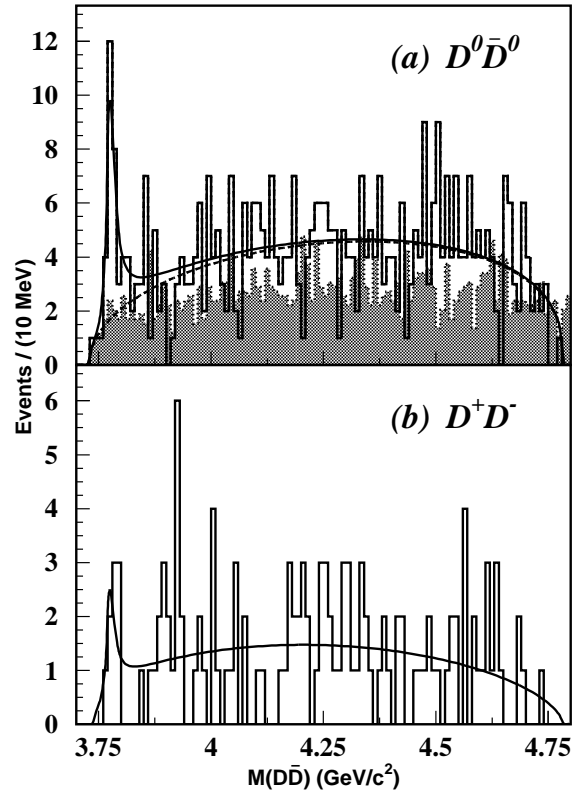}
%DRM \caption{ a): The $M(\dd)$ distribution for the events from the B-signal region.
%DRM Dashed line represents the background parameterization( see the text ).
%DRM Hatched histrogramm is constructed from the $\Delta E$ sidebands.  
\caption{ a): The $M(\dd)$ distribution for the events from the $B$-signal region. 
The dashed line represents the background parameterization (see the text). 
The hatched histogram is constructed from the $\Delta E$ sidebands.
b): Fitted $M(D^+D^-)$ distribution.   }  
\label{fitA2}
\end{figure}

%SLO The efficiencies for $B^+\ra\psi(3770)K^+$ followed by $\psi(3770)\ra\dd$ and 
% $\psi(3770)\ra D^+D^-$ were found to be $10.3\times 10^{-4}$ and $5.7\times 10^{-4}$, respectively. 
The MC-determined efficiencies for $B^+\ra\psi(3770)K^+$ followed by $\psi(3770)\ra\dd$ and 
$\psi(3770)\ra D^+D^-$ are $10.3\times 10^{-4}$ and $5.7\times 10^{-4}$, respectively. 
%DRM It gives 
This gives 
${\cal B}(B^+\ra\psi(3770) K^+)\times{\cal B}(\psi(3770)\ra\dd)=(0.34\pm 0.08\pm 0.05)\times 10^{-3}$
and ${\cal B}(B^+\ra\psi(3770) K^+)\times{\cal B}(\psi(3770)\ra D^+D^-)=(0.14\pm 0.08\pm 0.02)\times 10^{-3}$,
%DRM (0.14\pm 0.08\pm 0.03)\times 10^{-3}$. 
%DRM The first error is statistical and the second one is systematic. 
where the first error is statistical and the second is systematic. 
%DRM The systematic error in addition to the mentioned above comes from the 
In addition to the sources already mentioned, the systematic error comes from  
varying the signal and background shapes in $M(D\bar D)$ fitting ($5\%$)
%DRM and $\pi^0$ reconstruction efficiency ($6\%$). 
and from varying the $\pi^0$ reconstruction efficiency ($6\%$). 
%DRM From these two measurements we obtain their ratio to be equal to 
From these two measurements we obtain the ratio  
${\cal B}(\psi(3770)\ra \dd)/{\cal B}(\psi(3770)\ra D^+D^-)=2.43\pm 1.50\pm 0.43$. 
Given the large errors, our measurement is consistent with the 
previous measurement of this ratio by the MARK III 
Collaboration of $1.36\pm 0.23\pm 0.14$~\cite{ratio_mark3}.

%\begin{figure}[bth]
%\centering
%\includegraphics[width=0.5\textwidth]{helicity_v54.eps}
%\caption{ a): Helicity distribution for $B^+\ra\psi(3770)K^+$ decay followed by $\psi(3770)\ra\dd$. 
%%DRM Points are data obtained from the fit to the $M(\dd)$ in each ${\rm cos}\Theta_{\psi(3770)}$ bin.  
%%DRM The histogram shows the results of the fit by the expected form from MC.
%The points with errors are obtained from the fits to the $M(\dd)$ data in each ${\rm cos}\Theta_{\psi(3770)}$ bin.  
%The histogram shows the expected results based on signal MC (see text). 
%b): Helicity distribution for $B^+\ra\psi(3770)K^+$ decay followed by $\psi(3770)\ra D^+D^-$ 
%%DRM for data taken from the region of $M(D^+D^-)<3800$~MeV/c$^2$.}  
%for data taken from the region $M(D^+D^-)<3800$~MeV/c$^2$.}  
%\label{helicity}
%\end{figure}

\begin{figure}[bth]
\centering
\includegraphics[width=0.5\textwidth]{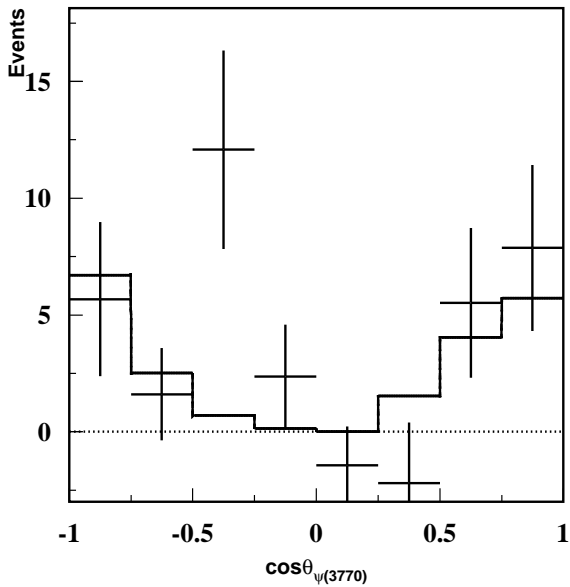}
\caption{ Helicity distribution for $B^+\ra\psi(3770)K^+$ decay followed by $\psi(3770)\ra\dd$. 
%DRM Points are data obtained from the fit to the $M(\dd)$ in each ${\rm cos}\theta_{\psi(3770)}$ bin.  
%DRM The histogram shows the results of the fit by the expected form from MC.
The points with errors are obtained from fits to the $M(\dd)$ data in each ${\rm cos}\theta_{\psi(3770)}$ bin.  
The histogram shows the expected distribution for $B^+\ra\psi(3770) K^+$  (see text). }
%b): Helicity distribution for $B^+\ra\psi(3770)K^+$ decay followed by $\psi(3770)\ra D^+D^-$ 
%%DRM for data taken from the region of $M(D^+D^-)<3800$~MeV/c$^2$.}  
%for data taken from the region $M(D^+D^-)<3800$~MeV/c$^2$.}  
\label{helicity}
\end{figure}

To extract ${\cal B}(B^+\ra\psi(3770) K^+)$ from the measurements of 
${\cal B}(B^+\ra\psi(3770) K^+)\times{\cal B}(\psi(3770)\ra \dd)$ 
and ${\cal B}(B^+\ra\psi(3770) K^+)\times{\cal B}(\psi(3770)\ra D^+D^-)$,  
%DRM we assume that $\dd$ and $D^+D^-$ modes totally saturate the $\psi(3770)$ 
we assume that the $\dd$ and $D^+D^-$ modes totally saturate the $\psi(3770)$ 
decay width. 
Summing both measurements gives 
${\cal B}(B^+\ra\psi(3770) K^+)=(0.48\pm 0.11\pm 0.07)\times 10^{-3}$. 

%Recently Belle observed strong signal for a new narrow charmonium state $X(3872)$ that decays to $J/\psi \pi^+\pi^-$~\cite{X3872}. 
%Since the mass of this new resonance is above  
%$D\bar{D}$ threshold it is important to get a 
%quantative information 
%for the rate of the decay $B^+\ra X(3872) K^+$ 
%followed by $X(3872)\ra D\bar{D}$. 
%It could help to 
%obtain additional information on its quantum numbers. 
%For this aim we fitted the $D^0\bar{D}^0$ and $D^+D^-$ invariant mass distributions 
%by including the possible contribution from $B^+\ra X(3872) K^+$, $X(3872)\ra D\bar{D}$ decays. 
%%The mass resolution was obtained from MC to be $..$~MeV . 
%Fit gives $2.5\pm 2.0$ and  $0.5\pm 1.3$ events for $D^0\bar{D}^0$ 
%and $D^+D^-$ cases respectively. 
%From this we have extracted a 90$\%$ CL upper limits of $7\times 10^{-5}$ 
%and $5\times 10^{-5}$ for 
%${\cal B}(B^+\ra X(3872) K^+)\times {\cal B}(X(3872)\ra D^0\bar{D}^0)$ 
%and ${\cal B}(B^+\ra X(3872) K^+)\times {\cal B}(X(3872)\ra D^+ D^-)$ respectively. 

Belle recently reported the observation of a narrow charmonium-like
state $X(3872)$ that decays to $\pi^+\pi^- J/\psi$~\cite{X3872}.  This state,
which is seen in the exclusive decay $B\ra K X(3872)$, is above
$D\bar{D}$ threshold. Information about the
$X(3872)\ra D\bar{D}$ decay rate would be useful for determining
its quantum numbers.  We refitted the $D^0\bar{D^0}$ and
$D^+D^-$ invariant mass distributions including possible
contributions from $B^+\ra X(3872)K^+,~X(3872)\ra D\bar{D}$
decays.  
%The mass resolution was obtained from MC to be $2$~MeV (see Fig.~\ref{x_mc}).
The fits yield $2.1\pm 1.8$ and $0.4\pm 0.8$ events
for the $D^0\bar{D^0}$ and $D^+D^-$ channels, respectively.
From this we determine $90\%~ CL$ upper limits
${\cal B}(B^+\ra X(3872)K^+)\times {\cal B}(X(3872)\ra D^0\bar{D^0})
<6\times 10^{-5}$
and
${\cal B}(B^+\ra X(3872)K^+)\times {\cal B}(X(3872)\ra D^+D^-)
<4\times 10^{-5}$.
We have also searched for a possible reflection from $B^+\ra X(3872) K^+$, $X(3872)\ra\dd\pi^0$ decays. 
This decay mode of the $X(3872)$ is interesting because it is predicted to be large if the $X(3872)$ 
is a $D\bar{D}^*$ multiquark ``molecular state''~\cite{voloshin_okun}. 
%We used the same selected sample of $\dd K^+$ combinations. 
A MC study shows that these decays produce a narrow, 
nearly gaussian reflection peak ($\sigma=9$~MeV) centered at $\Delta E=-145$~MeV.
%such a reflection become apparent as a narrow ($\sigma=9$~MeV) 
%gaussian peak centered at $\Delta E=-145$~MeV. 
Using the $D^0 \bar{D}^0 K^+$ signal described above, we require $M(\dd)$ 
to be less than $M(X(3872))-M(\pi^0)=3737$~MeV/c$^2$
%looked at the $\Delta E$ distribution 
%and fitted it to the gaussian with the fixed mean and $\sigma$ 
%to describe possible signal reflection and first order polynomial to describe the background. 
and fit the resulting $\Delta E$ distribution to a gaussian with mean and width
fixed at the values expected for the reflection
peak and a linear background contribution.
%a first-order polynomial to describe the background.
The fit 
%with the gaussian mean and $\sigma$ fixed from MC 
yields $2.2\pm 1.7$ events. 
From this we determine a $90\%~ CL$ upper limit 
${\cal B}(B^+\ra X(3872)K^+)\times {\cal B}(X(3872)\ra D^0\bar{D^0}\pi^0)
<6\times 10^{-5}$.

%DRM In summary, we performed the measurement of branching fraction for 
In summary, we have measured the branching fraction for 
$B^+\ra D^0 \bar D^0 K^+$ decay to be
${\cal B}(B^+\ra D^0 \bar D^0 K^+)=(1.17\pm 0.21\pm 0.15)\times 10^{-3}$. 
%DRM The search for $B^+\ra D^+D^-K^+$ decay results in the upper limit of 
A search for $B^+\ra D^+D^-K^+$ decay results in an upper limit of 
${\cal B}(B^+\ra D^+D^-K^+)<0.90\times 10^{-3} ~(90\%~ CL)$. 
%SLO By investigating the $\dd$ invariant mass spectrum from 
% $B^+\ra\dd K^+$ decays we observed a signal 
% %DRM near the mass of $3770$~GeV/c$^2$ with 
% %DRM the statistical significance $5.9\sigma$. 
% with a mass near $3770$~MeV/c$^2$ with 
% a statistical significance $5.9\sigma$, 
% %We identified it as the signal for the 
% which we attribute to
% exclusive $B^+\ra\psi(3770) K^+$ decay. 
We observe a peak in the $\dd$ invariant mass spectrum from 
$B^+\ra\dd K^+$ decays with a mass near $3770$~MeV/c$^2$  
that we attribute to exclusive $B^+\ra\psi(3770) K^+$ decay.
This signal, which has a  statistical significance of $5.9~\sigma$,
is the first observation of this decay mode. The mass of the $\psi(3770)$ 
is measured to be $3778.4\pm 3.0\pm 0.8$~MeV/c$^2$.
The value of ${\cal B}(B^+\ra\psi(3770) K^+)\times{\cal B}(\psi(3770)\ra\dd)$  
is measured to be $(0.34\pm 0.08\pm 0.05)\times 10^{-3}$. 
For $B^+\ra\psi(3770) K^+$ followed by $\psi(3770)\ra D^+D^-$ we extract
${\cal B}(B^+\ra\psi(3770) K^+)\times{\cal B}(\psi(3770)\ra D^+D^-)=(0.14\pm 0.08\pm 0.02)\times 10^{-3}$. 
%DRM (0.14\pm 0.08\pm 0.03)\times 10^{-3}$. The ratio of 
The ratio $\frac{{\cal B}(\psi(3770)\ra \dd)}{{\cal B}(\psi(3770)\ra D^+D^-)}$ 
%SLO was obtained to be $2.43\pm 1.50\pm 0.65$. 
is $2.43\pm 1.50\pm 0.43$. 
%DRM By assuming that $\dd$ and $D^+D^-$ modes totally saturate the $\psi(3770)$ 
By assuming that the $\dd$ and $D^+D^-$ modes totally saturate the $\psi(3770)$ 
decay width we obtain 
%DRM ${\cal B}(B^+\ra\psi(3770) K^+)=(0.48\pm 0.11\pm 0.12)\times 10^{-3}$ that 
${\cal B}(B^+\ra\psi(3770) K^+)=(0.48\pm 0.11\pm 0.07)\times 10^{-3}$ which 
is comparable to ${\cal B}(B^+\ra\psi(2S)K^+)$~\cite{pdg}. 
%DRM It gives an indication for a large $S$-$D$-mixing in $\psi(3770)$.  
This result suggests  a large amount of ${\rm S}$-${\rm D}$ mixing in the $\psi(3770)$. 

   For the decays $B^+\ra X(3872) K^+$ followed by $X(3872)\ra D^0\bar{D}^0$ and $D^+D^-$ 
we have set $90\%~ CL$ upper limits on ${\cal B}(B^+\ra X(3872) K^+)\times{\cal B}(X(3872)\ra D\bar{D})$ 
of $6\times 10^{-5}$ and $4\times 10^{-5}$ respectively. 
For the decay $B^+\ra X(3872) K^+$ followed by $X(3872)\ra D^0\bar{D}^0\pi^0$ we have set 
a $90\%~ CL$ upper limit 
of $6\times 10^{-5}$.

%\section*{Acknowledgments} 

%We wish to thank the KEKB accelerator group for the excellent
%operation of the KEKB accelerator.
%We acknowledge support from the Ministry of Education,
%Culture, Sports, Science, and Technology of Japan
%and the Japan Society for the Promotion of Science;
%the Australian Research Council
%and the Australian Department of Industry, Science and Resources;
%the National Science Foundation of China under contract No.~10175071;
%the Department of Science and Technology of India;
%the BK21 program of the Ministry of Education of Korea
%and the CHEP SRC program of the Korea Science and Engineering Foundation;
%the Polish State Committee for Scientific Research
%under contract No.~2P03B 01324;
%the Ministry of Science and Technology of the Russian Federation;
%the Ministry of Education, Science and Sport of the Republic of Slovenia;
%the National Science Council and the Ministry of Education of Taiwan;
%and the U.S.\ Department of Energy.

We wish to thank the KEKB accelerator group for the excellent
operation of the KEKB accelerator.
We acknowledge support from the Ministry of Education,
Culture, Sports, Science, and Technology of Japan
and the Japan Society for the Promotion of Science;
the Australian Research Council
and the Australian Department of Education, Science and Training;
the National Science Foundation of China under contract No.~10175071;
the Department of Science and Technology of India;
the BK21 program of the Ministry of Education of Korea
and the CHEP SRC program of the Korea Science and Engineering Foundation;
the Polish State Committee for Scientific Research
under contract No.~2P03B 01324;
the Ministry of Science and Technology of the Russian Federation;
the Ministry of Education, Science and Sport of the Republic of Slovenia;
the National Science Council and the Ministry of Education of Taiwan;
and the U.S.\ Department of Energy.


\begin{thebibliography}{99}
\bibitem{cc} References in this paper to a specific charged state are to be 
interpreted as implying the charge-conjugate state also.
\bibitem{stat_signif} The statistical significance is defined as $\sqrt{-2{\rm ln}(L_0/L{max})}$, where $L_0$ and $L_{max}$ are the likelihood with signal fixed at zero and at the fitted value, respectively.
\bibitem{de_sdb} The definition of the $\Delta E$ sidebands is $-0.080$~GeV $<\Delta E<-0.040$~GeV or $0.040$~GeV $<\Delta E<0.200$~GeV. 
%In case of multiple $B$ candidates we choose the candidate with the minimal value of 
If there is more than one B candidate, we choose the one with the smallest value of
$((M_{bc}-M_{B^+})/\sigma_{M_{bc}})^2$.
\bibitem{explanations} The mass resolution in the $\psi(3770)$ region was determined from MC to be 1~MeV/c$^2$.
\bibitem{def_helicity} The helicity angle $\theta_{\psi(3770)}$ is defined as 
the angle between the $D^0$ and $B^+$ momenta both calculated in the $\psi(3770)$ rest frame.
\end{thebibliography}
\end{document}